\documentclass[%
 reprint,superscriptaddress,
 amsmath,amssymb,
pra,
]{revtex4-2}

\usepackage{graphicx}
\usepackage{dcolumn}
\usepackage{bm}
\usepackage{float}

\begin{document}

\preprint{APS/123-QED}

\title{Development of a compact high-resolution absolute gravity gradiometer based on atom interferometers}

\author{Wei Lyu}
 \affiliation{State Key Laboratory of Magnetic Resonance and Atomic and Molecular Physics, Innovation Academy for Precision Measurement Science and Technology, Chinese Academy of Sciences, Wuhan 430071, China}
 \affiliation{School of Physical Sciences, University of Chinese Academy of Sciences, Beijing 100049, China}
\author{Jia-Qi Zhong}
 \email{jqzhong@apm.ac.cn}
\author{Xiao-Wei Zhang}
 \affiliation{State Key Laboratory of Magnetic Resonance and Atomic and Molecular Physics, Innovation Academy for Precision Measurement Science and Technology, Chinese Academy of Sciences, Wuhan 430071, China}
\author{Wu Liu}
 \affiliation{State Key Laboratory of Magnetic Resonance and Atomic and Molecular Physics, Innovation Academy for Precision Measurement Science and Technology, Chinese Academy of Sciences, Wuhan 430071, China}
 \affiliation{School of Physical Sciences, University of Chinese Academy of Sciences, Beijing 100049, China}
\author{Lei Zhu}
 \affiliation{State Key Laboratory of Magnetic Resonance and Atomic and Molecular Physics, Innovation Academy for Precision Measurement Science and Technology, Chinese Academy of Sciences, Wuhan 430071, China}
\author{Wei-Hao Xu}
 \affiliation{State Key Laboratory of Magnetic Resonance and Atomic and Molecular Physics, Innovation Academy for Precision Measurement Science and Technology, Chinese Academy of Sciences, Wuhan 430071, China}
 \affiliation{School of Physical Sciences, University of Chinese Academy of Sciences, Beijing 100049, China}
\author{Xi Chen}
\author{Biao Tang}
 \affiliation{State Key Laboratory of Magnetic Resonance and Atomic and Molecular Physics, Innovation Academy for Precision Measurement Science and Technology, Chinese Academy of Sciences, Wuhan 430071, China}
\author{Jin Wang}
\email{wangjin@apm.ac.cn}
\author{Ming-Sheng Zhan}
\email{mszhan@apm.ac.cn}
 \affiliation{State Key Laboratory of Magnetic Resonance and Atomic and Molecular Physics, Innovation Academy for Precision Measurement Science and Technology, Chinese Academy of Sciences, Wuhan 430071, China}
 \affiliation{Wuhan Institute of Quantum Technology, Wuhan 430206, China}

\date{\today}

\begin{abstract}
We present a compact high-resolution gravity gradiometer based on dual Rb-85 atom interferometers using stimulated Raman transitions. A baseline $L$=44.5 cm and an interrogation time $T$=130 ms are realized in a sensor head with volume of less than 100 liters. Experimental parameters are optimized to improve the short-term sensitivity while a rejection algorithm relying on inversion of the Raman wave vector is implemented to improve the long-term stability. After an averaging time of 17000 s, a phase resolution of 104 $\mu$rad is achieved, which corresponds to a gravity gradient resolution of 0.86 E. As far as we know, this is the sub-E atom gravity gradiometer with the highest level of compactness to date. After the evaluation and correction of system errors induced by light shift, residual Zeeman shift, Coriolis effect and self-attraction effect, the instrument serves as an absolute gravity gradiometer and with it the local gravity gradient is measured to be 3114 (53) E.

\end{abstract}

\maketitle


\section{\label{sec:level1}Introduction}

Gravimetry techniques play significant roles in the fields of metrology, geology, geophysics, seismology and industrial-related applications such as resource exploration and autonomous navigation. Since the inception in 1991 \cite{1}, the light-pulse atom interferometer (LPAI) has aroused great interest in gravity measurements and demonstrated remarkable resolution and long-term stability \cite{2,3,4,5,6,7}. After three decades of development, gravimeters based on LPAIs achieve such high maturity that six LPAI gravimeters have taken part in the 10th International Comparison of Absolute Gravimeters (ICAG) in 2017 and shown performances competing with classic corner cube gravimeters \cite{8,9,10}. Marine and airborne LPAI gravimeters have also been developed and showed performance beyond spring gravimeters \cite{11,12}. 

\begin{figure*}[t]
\centering
\includegraphics[width=150mm]{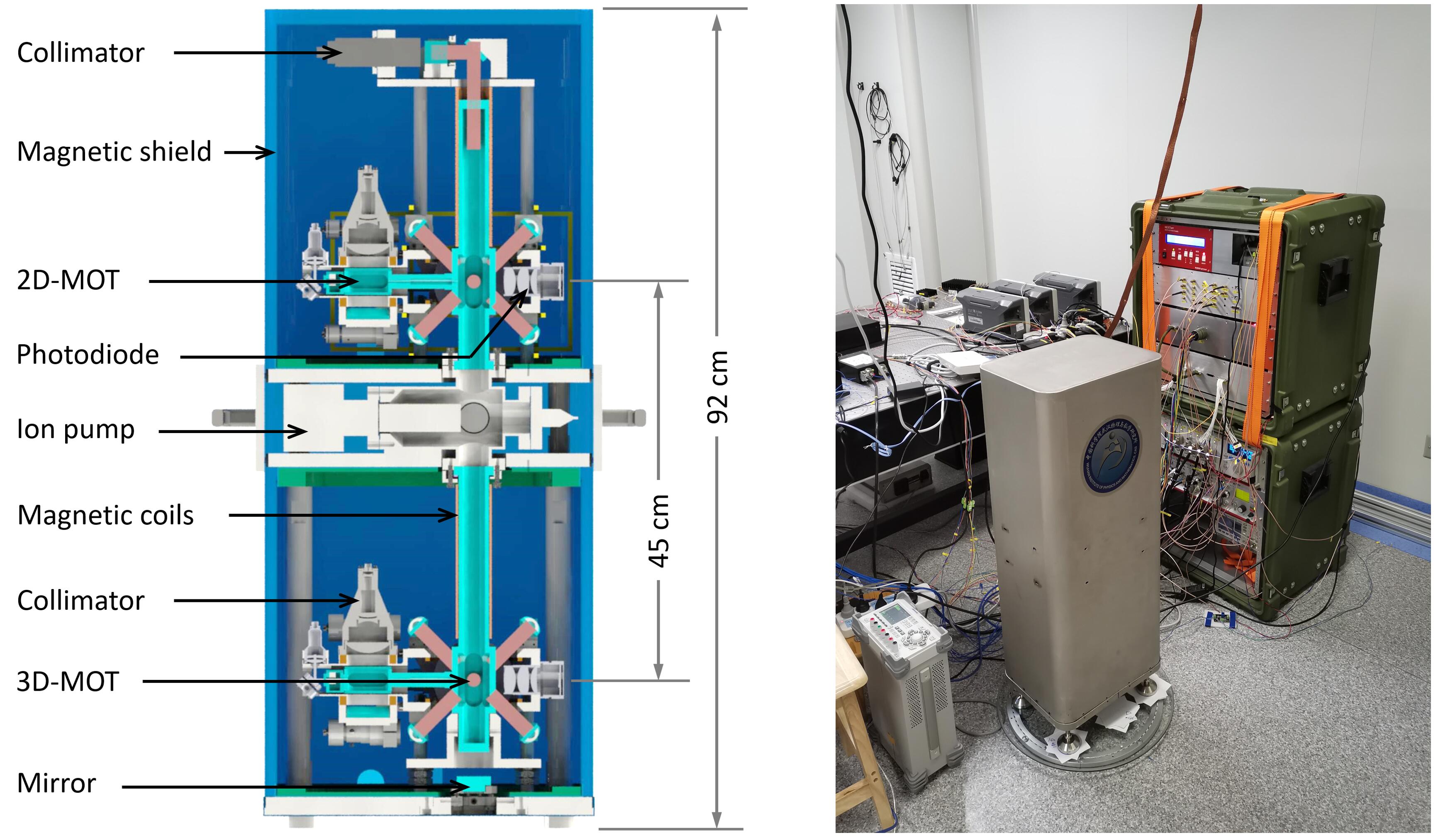}
\caption{\label{1} Design drawing and physical photo of the atom gravity gradiometer}
\end{figure*}

In comparison with gravimeters, gravity gradiometers could map subsurface density variations with higher resolutions. And due to the differential measurement mode, gravity gradiometers are highly immune from acceleration noises, that makes them useful in situations where vibration noise hampers gravity measurements. In LPAI based gravity gradiometers \cite{13}, simultaneous atom interferometers are manipulated by a common series of Raman laser pulses, which bring additional advantage of common mode rejection for many internal noises. However, among literatures, more works are focused on using the LPAI gravity gradiometers to measure the Newtonian gravitational constant \cite{14,15,16,17,18}, where the gradiometers serve as laboratory scientific installations, and only pay heed to the accuracy of relative changes of the gravity gradient produced by moving masses. In contrast, only a few research works have been reported about compact LPAI gravity gradiometers \cite{19} and absolute gravity gradient \cite{20}. Recently, the gravity gradiometer  in iXblue has reached an unprecedented resolution \cite{21}, that makes it more significant to make up for the weaknesses of atom gravity gradiometry in terms of compactness and accuracy. \\
\indent In this article, we present a compact and high-resolution gravity gradiometer based on dual Rb-85 LPAIs for the measurement of vertical gradient component. To achieve a longer baseline, a higher data rate, and more interrogation time within a smaller sensor head, two novel all-glass vacuum chambers with 2D magneto-optical traps (MOTs) are fabricated and the final detection process is carried out in the region of initial laser trapping. To save the measurement time and improve the sensitivity, the loading time of the MOTs is suppressed thus make the interferometers operating at the detection noise limit. To improve the long-term stability, an algorithm relying on inversion of the Raman wave vector k is implemented and the k-polarity-independent drift is significantly removed. The improved short-term sensitivity and long-term stability lead to a measurement resolution up to 0.86 E after an integration of 17000 s. Finally, the measurement error for absolute gravity gradient is elaborately studied. Besides those caused by Zeeman shifts, self-attraction and Coriolis effect, a special measurement error caused by the Raman laser scheme based on chirping acousto-optical modulator (AOM) is also evaluated.

\section{\label{sec:level1}Principle and apparatus}

Our gradiometer is based on two atom interferometers (gravimeters) apart in the vertical direction to measure the differential acceleration between two clouds of cold rubidium atoms in free fall. In each gravimeter, the cold atom cloud is split, redirected, and recombined by three Raman laser pulses, which is analogous to the beam splitter and mirrors in optical M-Z interferometers, then the excited atoms are imprinted by the phase of Raman laser pulses. Since the gravitational acceleration g can be evaluated by three positions of the mass ($z_1$, $z_2$, $z_3$) and the time interval $T$, 
\begin{equation}
g=(z_1-2z_2+z_3)T^2,
\end{equation}
the phase shift of atom interferometer induced by $\bm{g}$ can be written as
\begin{equation}
\Delta\phi=\phi_1-2\phi_2+\phi_3=\bm{k}_\mathrm{eff}\cdot\bm{g}T^2,
\end{equation}
where $\phi_i =\bm{k}_\mathrm{eff}\cdot\bm{z}_i,(i=1, 2, 3)$ are the laser phases.

For two gravimeters A and B separated by $\bm{L}$, if the atoms are all initially prepared in one of the two ground states $|1\rangle$ and $|2\rangle$, the probability of finding the atoms in the other state after the interference is
\begin{subequations}
\begin{equation}
P_\mathrm{A}=1/2(C_\mathrm{A}-D_\mathrm{A}\cos(\phi_0+\Delta\varphi_\mathrm{A})),
\end{equation}
\begin{equation}
P_\mathrm{B}=1/2(C_\mathrm{B}-D_\mathrm{B}\cos(\phi_0+\Delta\varphi_\mathrm{B})),
\end{equation}
\end{subequations}
where $C_A$, $C_B$ and $D_A$, $D_B$ are parameters that denotes the offset and contrast of the two interferometer signals respectively. And the phase difference is
\begin{equation}
\Delta\varphi_\mathrm{diff}=\Delta\varphi_\mathrm{A}-\Delta\varphi_\mathrm{B}=\bm{k}_\mathrm{eff}\cdot\Delta\bm{g}T^2,
\end{equation}
where $\Delta\bm{g}=\bm{g}_\mathrm{A}- \bm{g}_\mathrm{\emph{B}}$ is the acceleration difference between the two gravimeters. Then the gravity gradient can be calculated by
\begin{equation}
\mit{\Gamma}=\mathrm{\Delta}\varphi_\mathrm{diff}/(\textbf{\emph{k}}_\mathrm{eff}\cdot\textbf{\emph{L}}T^2),
\end{equation}
In most cases, for gravity gradient measurement, we only care about the acceleration difference rather than the exact accelerations at either position of the two gravimeters. Therefore, usually we use the least-squares fitting of the ellipse algebraic formula \cite{22}
\begin{equation}
aP_\mathrm{A}^2+bP_\mathrm{A}P_\mathrm{B}+cP_\mathrm{B}^2+dP_\mathrm{A}+eP_\mathrm{B}+f=0
\end{equation}
to evaluate the parameters $(a, b, c, d, e, f)$ and to obtain the differential phase
\begin{equation}
\Delta\varphi_\mathrm{diff}=\cos^{-1}(-b/(2\sqrt{ac})).
\end{equation}

\begin{figure*}[t]
\centering
\includegraphics[width=130mm]{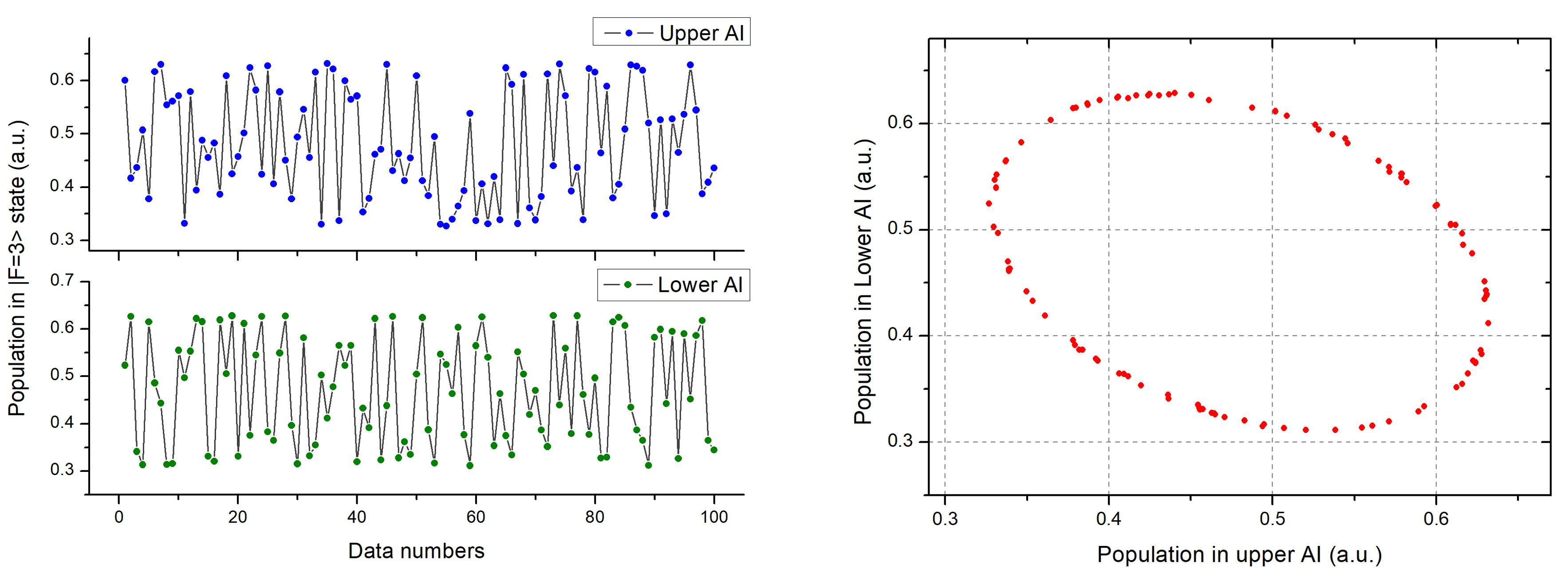}
\caption{\label{2} Interferometer signals of the two AIs in discrete and differential modes. The left and right graphs are from a same group of data with 100 measurements.}
\end{figure*}

Experimentally, in each gravimeter, Rb-85 atoms are laterally precooled by 2D-MOT and transferred to 3D-MOT for further cooling and trapping. Cold atom clouds with temperature of about 2.8 $\mu$K are then launched upwards at a velocity of about 1.8 m/s by moving optical molasses technique. During the flight, a bias magnetic field is applied to define the quantization direction, and atoms are firstly selected to magnetically insensitive state with the magnetic quantum number $m_F=0$. Then Raman laser pulse sequence $\pi/2-\pi-\pi/2$ is applied. The first pulse with the duration $\tau_1=\pi/(2\Omega)=4$ $\mu$s acts as a splitter, where $\Omega$ is the two-photon Rabi frequency, which drives the atom into an equal superposition of the two ground states $F=3$ and $F=2$. The second pulse with the duration $\tau_2=\pi/\Omega=8$ $\mu$s acts as a mirror, which exchanges the populations in the two states. The last pulse with the duration $\tau_3=\tau_1$ recombines the atom wave packets and complete the interference process. At the end of a measurement circle, populations in both ground state are detected successively. To remove the fluctuation of the total atom numbers, least-squares fitting of the ellipse algebraic formula (6) is carried out with normalized populations
\begin{subequations}
\begin{equation}
P\mathrm{_A^{nom}}=P\mathrm{_A}^{F=3}/(P\mathrm{_A}^{F=3}+\eta_\mathrm{A}P\mathrm{_A}^{F=2}),
\end{equation}
\begin{equation}
P\mathrm{_B^{nom}}=P\mathrm{_B}^{F=3}/(P\mathrm{_B}^{F=3}+\eta_\mathrm{B}P\mathrm{_B}^{F=2}),
\end{equation}
\end{subequations}
where $\eta_\mathrm{A}$ and $\eta_\mathrm{B}$ are the relative detection efficiencies for state $|F=3\rangle$ and $|F=2\rangle$ in the two interferometers respectively. In our gradiometer, the two gravimeters have their own detection system, which makes  $\eta_\mathrm{A}$ and $\eta_\mathrm{B}$ a little different.

The design and the physical photo of the sensor head are shown in Fig.\ref{1}. To achieve a higher data rate, two identical atom fountain setups are employed, each of them has its own 2D and 3D-MOT, which enable the two cold atom clouds and fountains be prepared simultaneously. Compared to the scheme of juggling \cite{15,16}, this scheme allows us to realize a relatively high sampling rate up to 1.7 Hz at expense of simplicity of the sensor head and laser system. To achieve a higher grade of noise common mode rejection, the vacuum chambers of the two setups are connected as one. The Raman laser delivered by a polarization-maintaining optical fiber is collimated into a 20 cm diameter beam, which vertically penetrates the vacuum chamber from the top window, and is retroreflected by a mirror under the bottom window. Since there is neither window nor air between the two atom clouds in the laser path, in comparison with the scheme based on two independent vacuum chambers, the phase drifts induced by the windows’ beam steering effect and the air’s refraction index variations are eliminated \cite{17}. 

Although the physical structure is similar to our former generation \cite{23}, many important improvements have been implemented. To squeeze the dimensions of the sensor head, except for the vacuum pumping unit, the vacuum chamber is entirely fabricated by JGS1 silica. Due to the active nature of alkali metals, the sample source is the most troublesome part for glass vacuum chambers. In previous glass chambers, the reservoir of alkali metal is still a piece of soft metal or connected with a copper transmission pipe that cannot be removed \cite{19,24}. Both solutions include a series of metal screws and accessories, which makes the vacuum chamber complicated and malformed. In our newly developed vacuum chamber, we freeze the rubidium samples to -200 ℃ with liquid nitrogen to reduce the activity, then put the frozen samples directly to the silica reservoir and seal the chamber within a few minutes. By this new method, two compact real glass rubidium sample sources are realized, which enables the optical and magnetic accessories for each 2D-MOT be integrated as a single module and directly sheathed to the 2D-MOT chamber from the terminal side.

Thanks to the large aperture of the glass chamber and the elaborate design of interior space layout, with a 44.5 cm baseline and two 18 cm fountains, the height of the total sensor head including two-layer magnetic shields is only 92 cm. To make full use of the vertical space of the sensor head, the final detection process is carried out in the region of initial 3D-MOT. Since the 3D-MOT is on the axis of the 2D-MOT, this design brings the trouble of a strong detection background. During the detection phase, many atoms from the 2D-MOT with low axial velocity arrive at the detection region hundreds of milliseconds after the moving molasses, which produces a strong background whose amplitude is even higher than that from the coherent atoms. To solve this problem, we perform a horizontal blow away before the atoms fall back to the detection region with the horizontal pushing beam, which have transported atoms from 2D-MOT to 3D-MOT at the laser trapping stage. By this operation, most of the atoms in the horizontal path are removed, and the background of final detection signal is reduced by almost 90\%. 

A detailed description of the laser system can be found in \cite{25}. Two diode lasers and two tapered amplifiers are employed in the system. The main laser is stabilized on the transition of $|5^2S_{1/2},F=3\rangle\rightarrow|5^2P_{3/2},F´=4\rangle$ and amplified by one amplifier, then serves for cooling, fountain, blow-away, detection and Raman laser seeding. The other laser is stabilized on the transition $|5^2S_{12/},F=2\rangle\rightarrow|5^2P_{3/2},F'=3\rangle$ and serves as repumping laser. The Raman laser is sourced from the amplified main laser and generated by a 1.5 GHz AOM. The +1 and -1 order diffraction beams are recombined and injected into the other tapered amplifiers; thus, a pair of Raman beams with a frequency difference equal to the $5^2S_{1/2}$ hyperfine splitting is obtained. In comparison with the method of phase locking and electro-optic modulation, this scheme avoids the risk of out of lock and complicated processing for the multi-sideband effects, resulting in higher robustness and measurement stability. However, this configuration could also bring an imbalanced ac-Stark shift during the frequency chirping, and results in a non-negligible measurement error. In this work, a real-time feedback control of the intensity ratio is performed to eliminate this effect, which we describe in more detail in Sec.V.

There are no vibration isolation or compensation devices in the gravity gradiometer, so each interferometer suffers from vibration noise. In general laboratory environment, no fringe can be read from either individual output signals due to the sensitivity of single AI to vibration. However, when operating in differential measurement mode, as shown in Fig.\ref{2}., they exhibit very strong immunity for vibration noise. We have moved the gradiometer to a quiet underground environment, where the two interferometers could output clear fringes and serve as gravimeters, but found no difference about the level of ellipse phase noise. 

\section{\label{sec:level1}Sensitivity}

For a quantum system with white noise, the measurement resolution R can be expressed as
\begin{equation}
R=S/\sqrt{\tau},
\end{equation}

where $S$ is the sensitivity and $\tau$ is the measurement time. An improved measurement resolution R could be achieved by increasing the sensitivity $S$ or extending the measurement time $\tau$. And for atom interferometers based on discrete atom clouds, the sensitivity $S$ is
\begin{equation}
S=N_s\sqrt{\delta t},
\end{equation}
where $N_s$ is the single shot noise and $\delta t$ is the single measurement time. Therefore, both suppression of the single shot noise $N_s$ and reduction of the single measurement time $\delta t$ help to improve the sensitivity.

\begin{figure}[h]
\centering
\includegraphics[width=75mm]{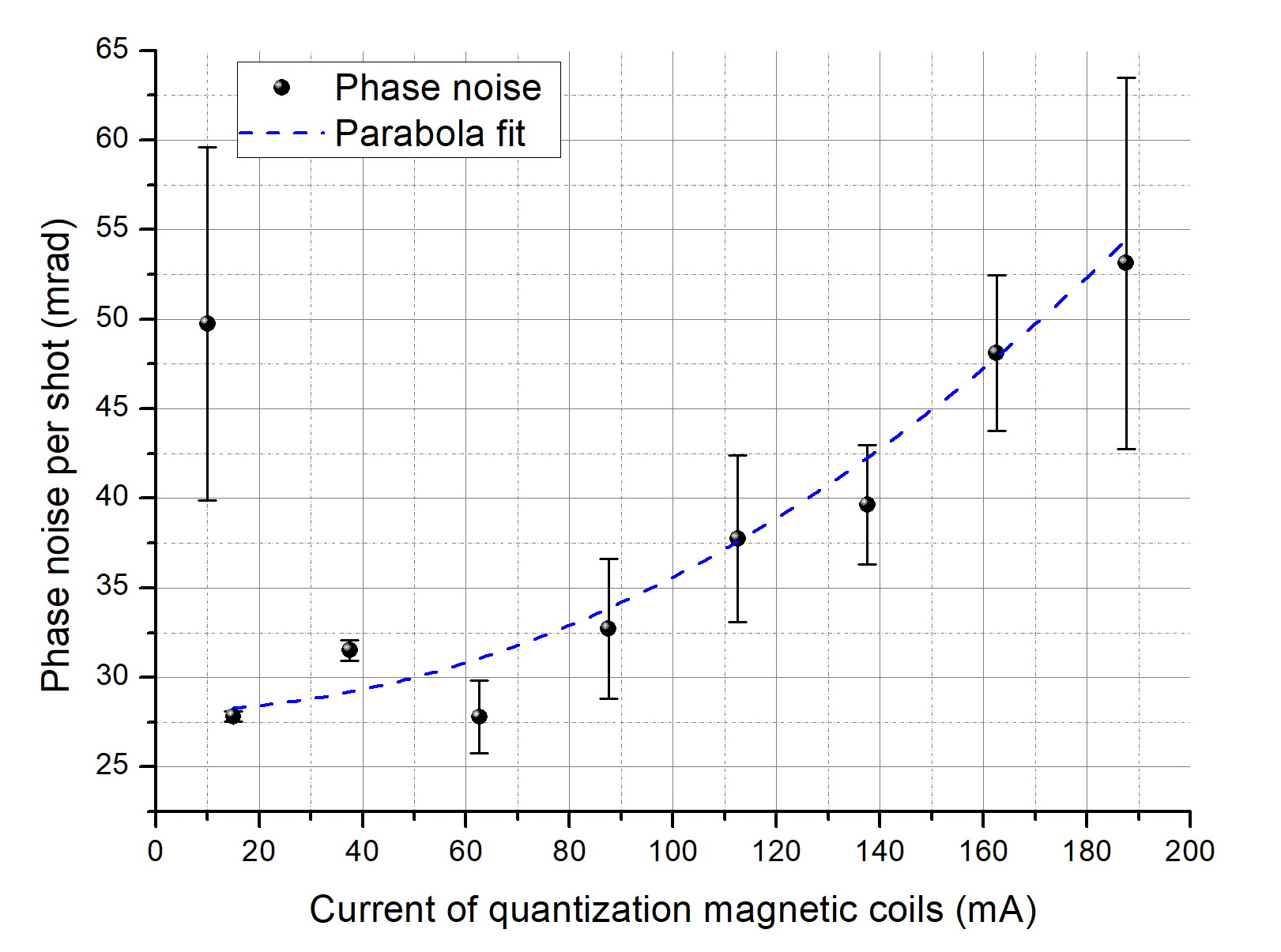}
\caption{\label{3} Dependence of single shot noise on the current of bias coils. Each point corresponds 3 measurements of 40 minutes. The dashed line is parabola fit of the data from the second point.}
\end{figure}

Many works have been reported to suppress the single shot noise, such as active control of the critical parameters \cite{26} and utilization of normalized detection \cite{27,28}. In this work, we investigate the dependence of single shot noise on the quantization magnetic field. As shown in Fig.\ref{3}., the ellipse phase noise has minimum values between 15$\sim$60 mA and increases significantly as the magnetic field increases due to the noise of quadratic Zeemanshift. When the current decreases to 10 mA, the phase noise increases drastically due to the loss of quantization direction. According to this result, we choose a quantization magnetic current of 15 mA in the following experiments, which corresponds to a magnetic field of about 30 mG.

\begin{figure}[h]
\centering
\includegraphics[width=73mm]{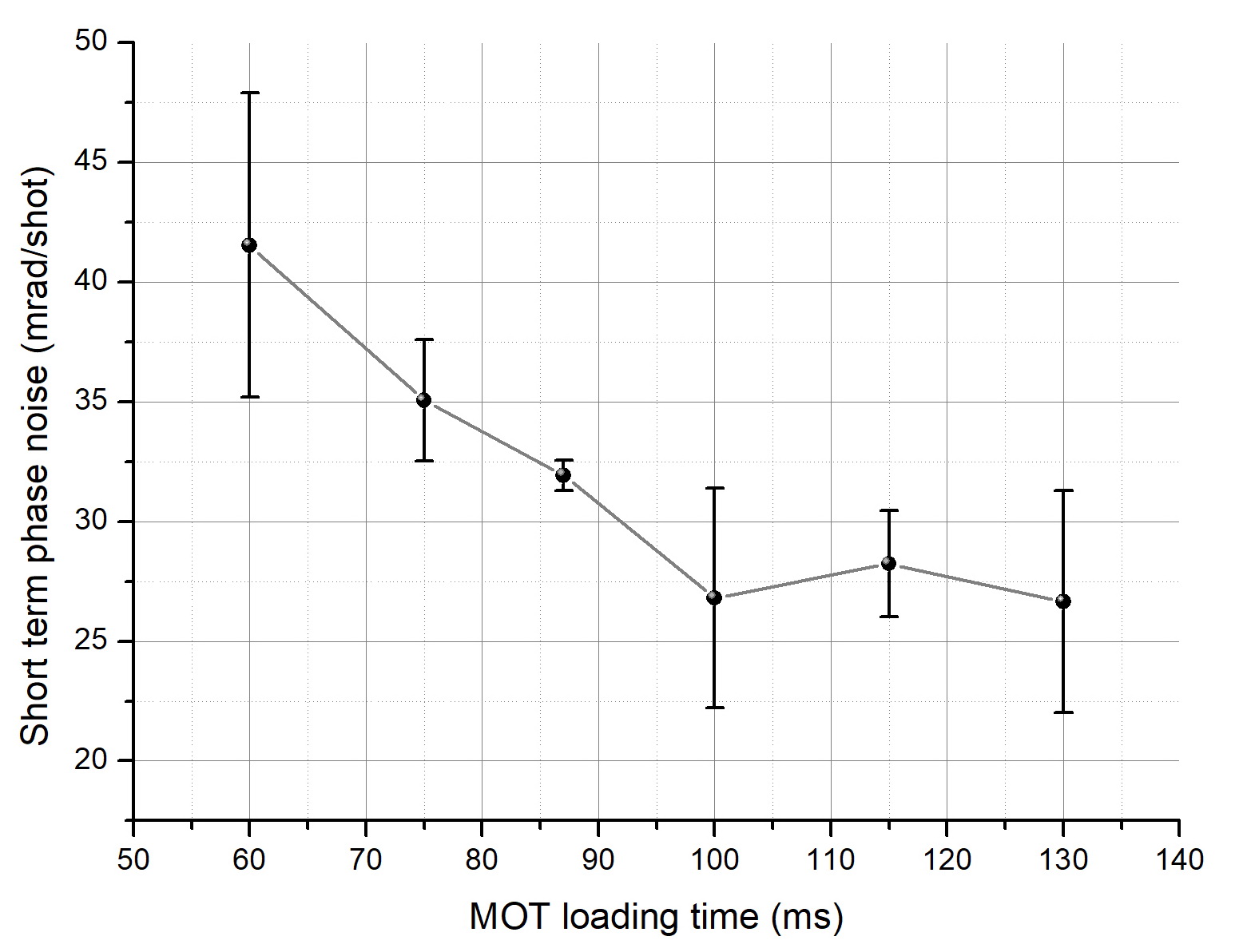}
\caption{\label{4} Dependence of single shot noise on the MOT loading time.}
\end{figure}

From Eq.(10), we know that with a constant single shot noise $N_s$, a higher data rate $1/\delta t$ also corresponds to a higher sensitivity $S$. Before the MOT is saturated, a longer loading time corresponds to a higher cold atom number $N$ and lower detection and quantum projection noise limits. However, as the loading time occupies a large part of the measurement time $\delta t$, for a system with other technical noises higher than the two noise limits, a large atom number $N$ is useless for the total noise level, and the redundant loading time, in turn, will decrease the sensitivity $S$. Here we investigate the relationship between the single shot noise and the MOT loading time to find an optimal loading time that yields the highest sensitivity. The result is shown in Fig.\ref{4}., as the loading time increases, the single shot noise decreases and reaches a nearly constant level. Both quantum projection and detection phase noises are inversely related to atom number, but the functional relationships are different. Given a detection noise with a constant amplitude, which comes from detector and detection lasers, the corresponding phase noise will show linear relationship with the absolute amplitude of interferometer signal, which is proportional to the atom number. In contrast, for the quantum projection noise, it is a well-known $1/\sqrt{N}$ relationship between the transferred phase noise and the atom number. Therefore, from Fig.\ref{4}., the noise from detection is more dominant than that from the quantum projection. Consequently, a loading time of 100 ms is chosen in the following experiments.

\begin{figure}[h]
\centering
\includegraphics[width=75mm]{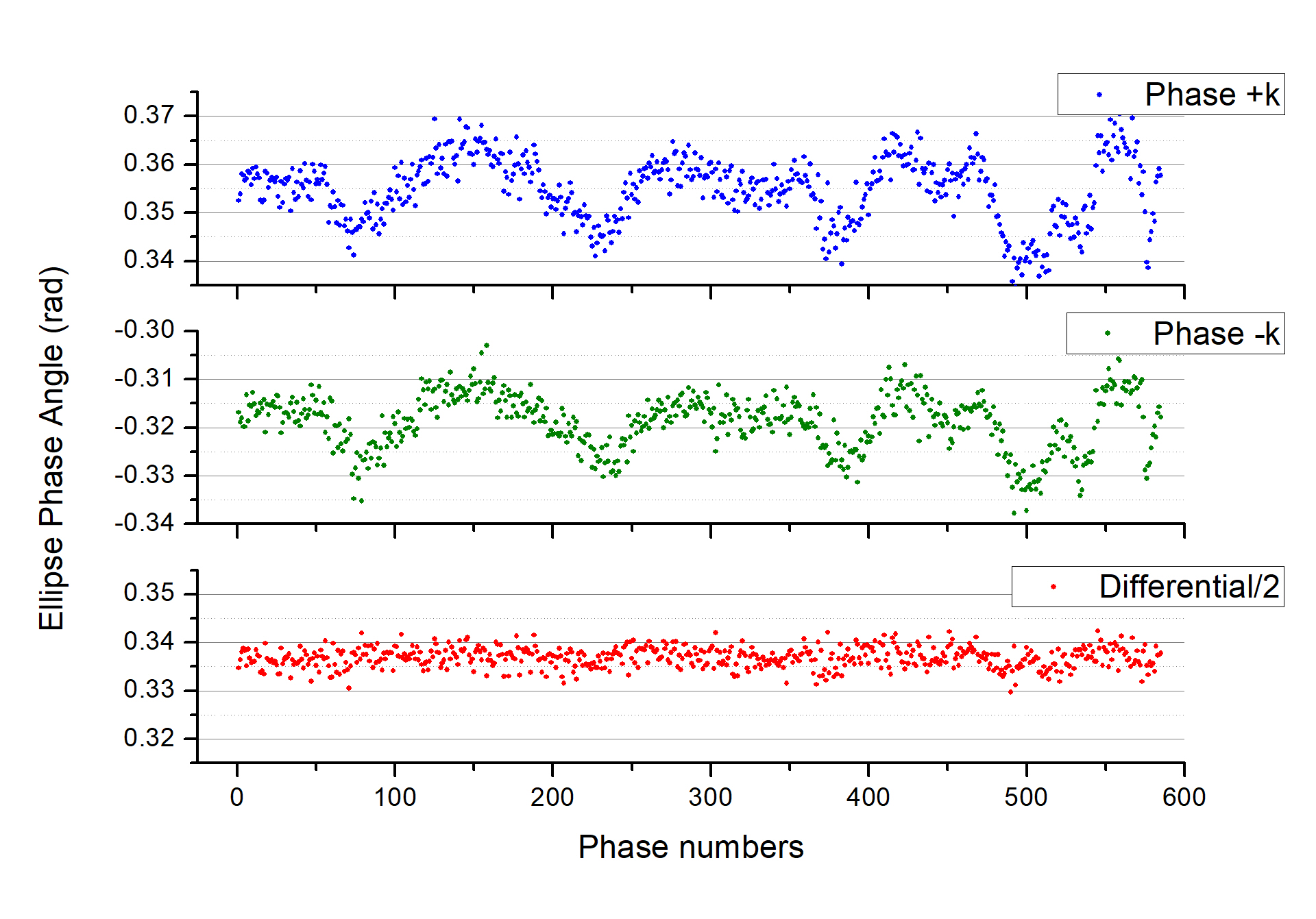}
\caption{\label{5} Improvement of the long-term stability. Three sets of data (from top to bottom) are the phases $\Delta\varphi_{+k}$, $\Delta\varphi_{-k}$ and $(\Delta\varphi_{+k}-\Delta\varphi_{-k})/2$ respectively, each data of $\Delta\varphi_{+k}$ and $\Delta\varphi_{-k}$ is extracted from 40 measurements.}
\end{figure}
\begin{figure}[h]
\centering
\includegraphics[width=80mm]{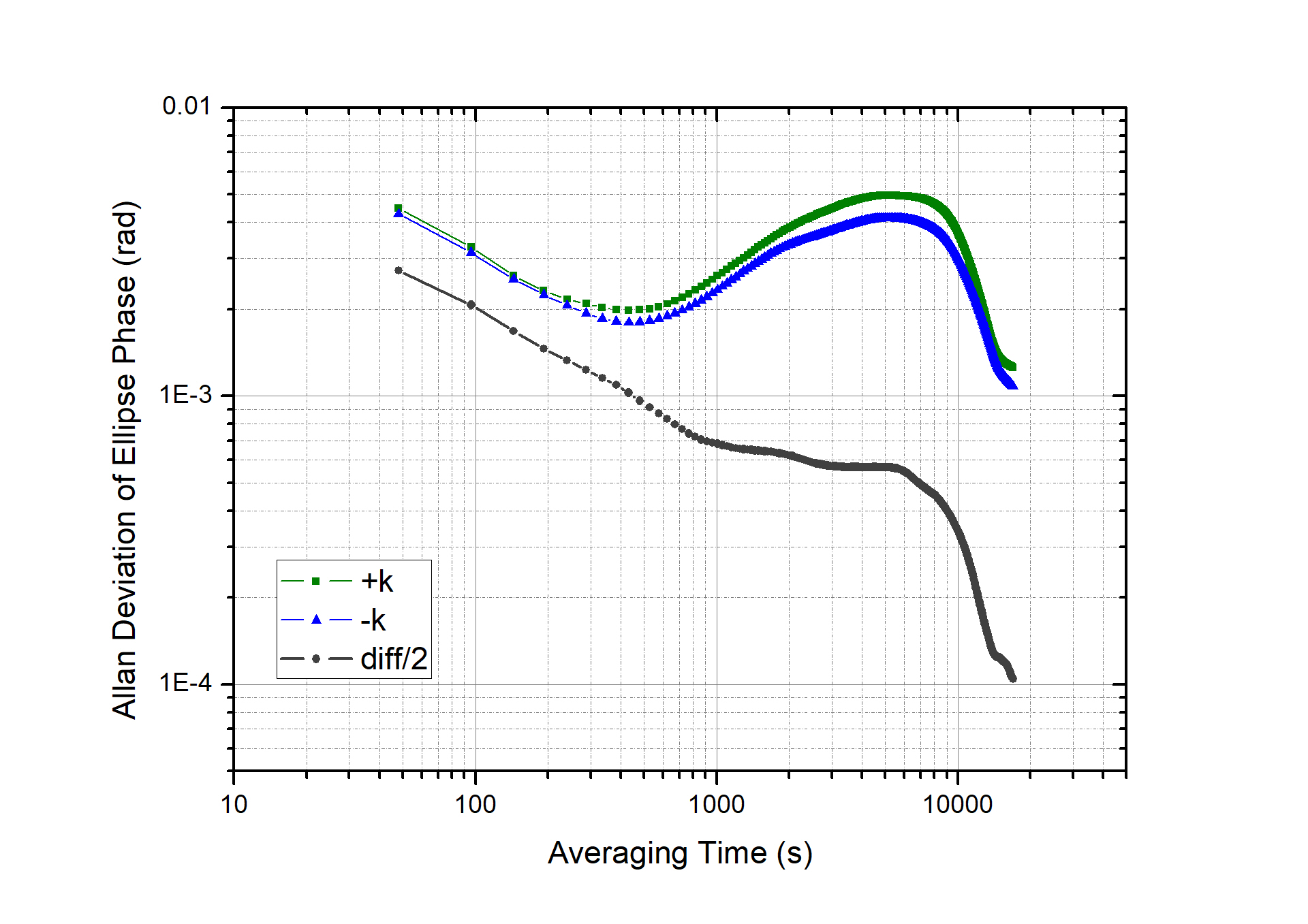}
\caption{\label{6} Allan deviations for the ellipse phases in Fig. 4. Each data pair $\Delta\varphi_{+k}$ and $\Delta\varphi_{-k}$ are extracted from 80 measurements and cost 48 s.}
\end{figure}

\section{\label{sec:level1}Long-term stability}

From Eq.(9), we know that in a white noise system, the measurement resolution is inversely proportional to the square root of the measurement time $\tau$. However, there is no technical noise which is completely white in AIs. Many parameters, such as laser frequencies, laser intensities and magnetic field strengths, vary with time and environmental factors, which induces long-term phase drift in AIs. Here we employ an algorithm relying on inversion of the Raman wave vector $\bm{k}$ to remove the k-polarity-independent drifts and improve the long-term stability of the gradiometer.
The differential phase between the two AIs can be written as 
\begin{equation}
\Delta\varphi=\Delta\varphi_{k\mathrm{\_ori\_depen}}+\Delta\varphi_{k\mathrm{\_ori\_indepen}},
\end{equation}
where $\Delta\varphi_{k\mathrm{\_ori\_depen}}$ is the phase whose polarity depends on the direction of Raman wave vector $\bm{k}$, it is usually induced by the gravity and rotation. $\Delta\varphi_{k\mathrm{\_ori\_indepen}}$ is the phase whose polarity is independent on the direction of $\bm{k}$, which is usually induced by Zeeman shift and light shift, etc. We alternately measure $\Delta\varphi$ with $+k$, whose vector points downwards, and $-k$, whose vector points upwards, and get
\begin{subequations}
\begin{equation}
\Delta\varphi_{+k}=\Delta\varphi_{k\mathrm{\_ori\_indepen}}+\Delta\varphi_{k\mathrm{\_ori\_depen}},
\end{equation}
\begin{equation}
\Delta\varphi_{-k}=\Delta\varphi_{k\mathrm{\_ori\_indepen}}-\Delta\varphi_{k\mathrm{\_ori\_depen}}.
\end{equation}
\end{subequations}

Then the phase $\Delta\varphi_{k\mathrm{\_ori\_depen}}$ including the effect of gravity gradient could be extracted as
\begin{equation}
\Delta\varphi_{k\mathrm{\_ori\_depen}}=\frac{1}{2}(\Delta\varphi_{+k}-\Delta\varphi_{-k}),
\end{equation}

We carried out a long-term interleaving measurement of $\Delta\varphi_{+k}$ and $\Delta\varphi_{-k}$ ,and the results are shown in Fig. \ref{5}. Apparent synchronous phase drifts happened at $\Delta\varphi_{+k}$ and $\Delta\varphi_{-k}$, however, the differential phase $(\Delta\varphi_{+k}-\Delta\varphi_{-k})/2$ shows a high stability. The Allan deviations are shown in Fig. \ref{6}. In comparison with the single measurement of $\Delta\varphi_{+k}$ and $\Delta\varphi_{-k}$, the long-term stability of the differential phase is improved by an order of magnitude and reaches a level of 104 $\mu$rad after an averaging time of 17000 s. For a gravity gradiometer with a baseline $L$=44.5 cm and an interrogation time $T$=130 ms, the corresponding resolution of gravity gradient is 0.86 E. The Allan deviations of the differential phase does not decrease along a constant slope, but deviates from the ideal trend from 1000 s to 14000 s, indicating that there are still residual factors with a specific period perturbing the measurement. This period is roughly in accordance with the temperature fluctuations in our laboratory, thus a temperature control will be carried out in our future works. 

\section{\label{sec:level1}Measurement errors}

For an absolute gravity gradiometer, accuracy is another important technical index in addition to the sensitivity and the long-term stability. Although gradiometers have the similar error sources with gravimeters, the roles played by each source are different. Due to the mechanism of common-mode rejection, many important errors in atom gravimeters, which relate to laser phases, such as those caused by laser path variation \cite{9} and synchronous vibrations \cite{4,9}, are negligible in gravity gradiometers. Additionally, unlike the gravimeters with targets to resolve n$g$ signals from a background of 1 $g$, for gradiometers to distinguish 1 E from $10^3$ E, the errors caused by the inaccuracy of the effective wave vector, originating from laser frequency, are also not necessary to be considered.

\begin{figure}[h]
\centering
\includegraphics[width=80mm]{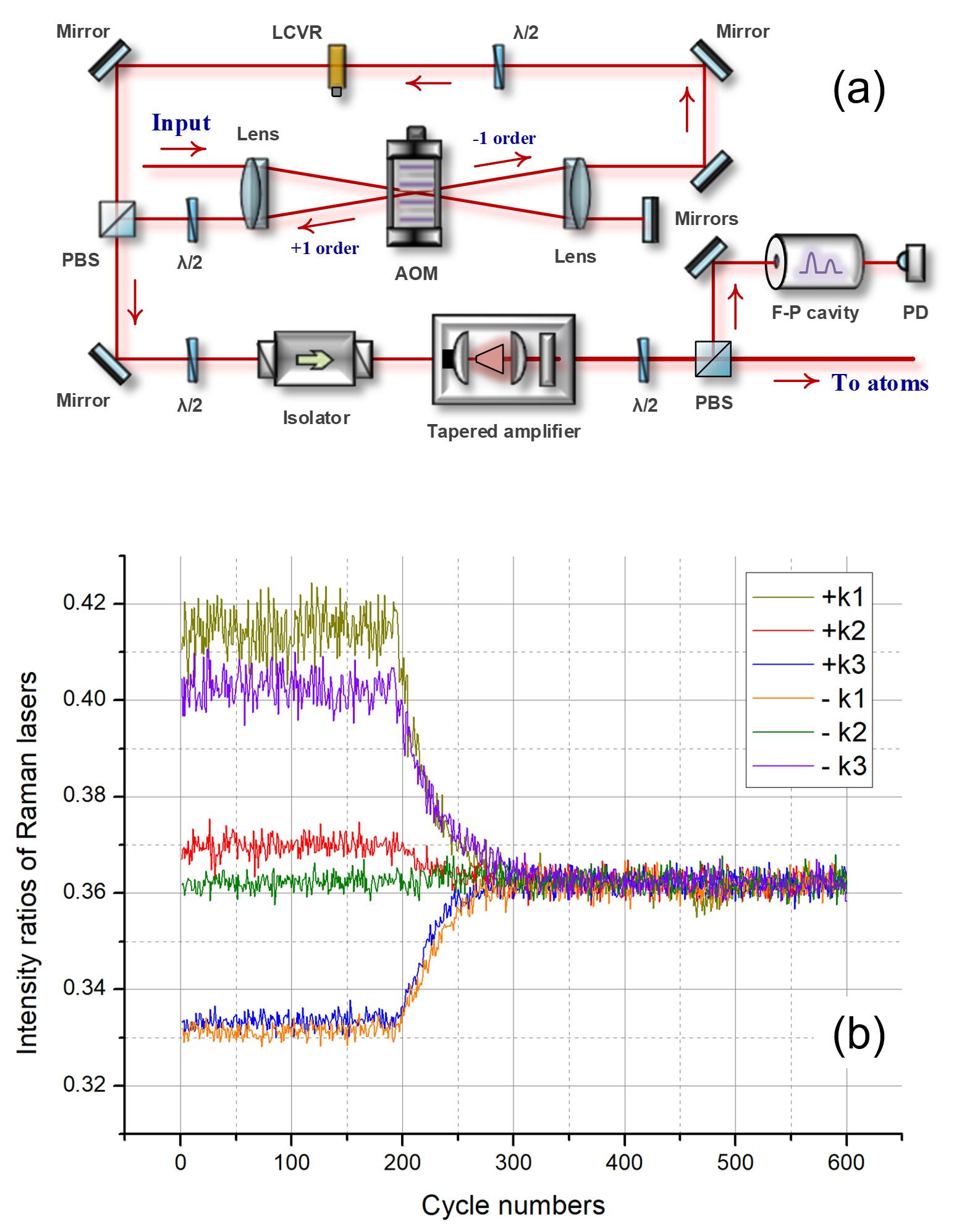}
\caption{\label{7} The schematic scheme of the Raman laser (a) and the effect of feedback control of the laser intensity ratios (b). The +1 and -1 order diffraction light of the AOM are combined via a PBS and amplified simultaneously by a tapered amplifier. A liquid crystal variable retarder (LCVR) is utilized to adjust the injecting intensity ratio of the two Raman beams.}
\end{figure}

In contrast, some insignificant errors in atom gravimeters, such as that induced by light shift, play more significant roles in our gradiometer. The differential ac Stark shift at each Raman laser pulse is given by
\begin{equation}
\delta^{ac}=\Omega_a^{ac}-\Omega_b^{ac}=\sum\limits_{m,i}(\frac{|\Omega_{a,i,m}|^2}{4\Delta_{a,i,m}}+\frac{|\Omega_{b,i,m}|^2}{4\Delta_{b,i,m}}),
\end{equation}
where $\Omega$ is one-photon Rabi frequency, $\Delta$ is detuning, $a$, $b$ and $i$ respectively denote the two ground states and the excited state in stimulated Raman transition, while $m$ is the index of light field. Usually, an optimal intensity ratio of the Raman laser components is set to cancel out the ac Stark shift $\delta^{ac}$ to zero. However, in our Raman laser scheme (shown in Fig. \ref{7}(a)), since both the diffraction efficiency of AOM and the amplification efficiency of TA vary with the modulation frequency, the intensity ratio changes from pulse to pulse when the frequency is chirping to compensate the gravity induced doppler shift. During a cycle consisting of a $+k$ and a $-k$ measurements, a total of six modulation frequencies are employed, corresponding to six intensity ratios. This leads to a phase shift
\begin{equation}
\Delta\varphi^{ac}=2(\delta_1^{ac}-\delta_3^{ac})\tau
\end{equation}
in each $+k$ or $-k$ measurement, where $\delta_1^{ac}$ and $\delta_3^{ac}$ are the ac Stark shifts at the first and the last Raman pulses respectively and cannot be canceled because all the intensity ratios are different from each other. Additionally, the projections of atom trajectories in two AIs on the horizontal plane cannot be exactly overlapped, that makes this light shifts different in the two AIs and the corresponding phase shifts cannot be completely rejected.

To solve this problem, we programmed a real-time feedback control of the intensity ratio between Raman laser components. A F-P cavity is employed to read the intensity ratio, and a liquid crystal variable retarder (LCVR) is used to perform the adjustment. The effect of the control loop is shown in Fig. \ref{7}(b). In free running mode, the intensity ratios distribute from 0.33 to 0.42, and when the feedback turns on, within 100 cycles, all the ratios are drawn to the set value of 0.362, which is the optimal intensity ratio determined by earlier experiment. With respect to the ellipse phase, an offset of +11.0±0.8 mrad is observed after the feedback control turns on, which corresponds to a gravity gradient of +91±6 E.

\begin{figure}[h]
\centering
\includegraphics[width=90mm]{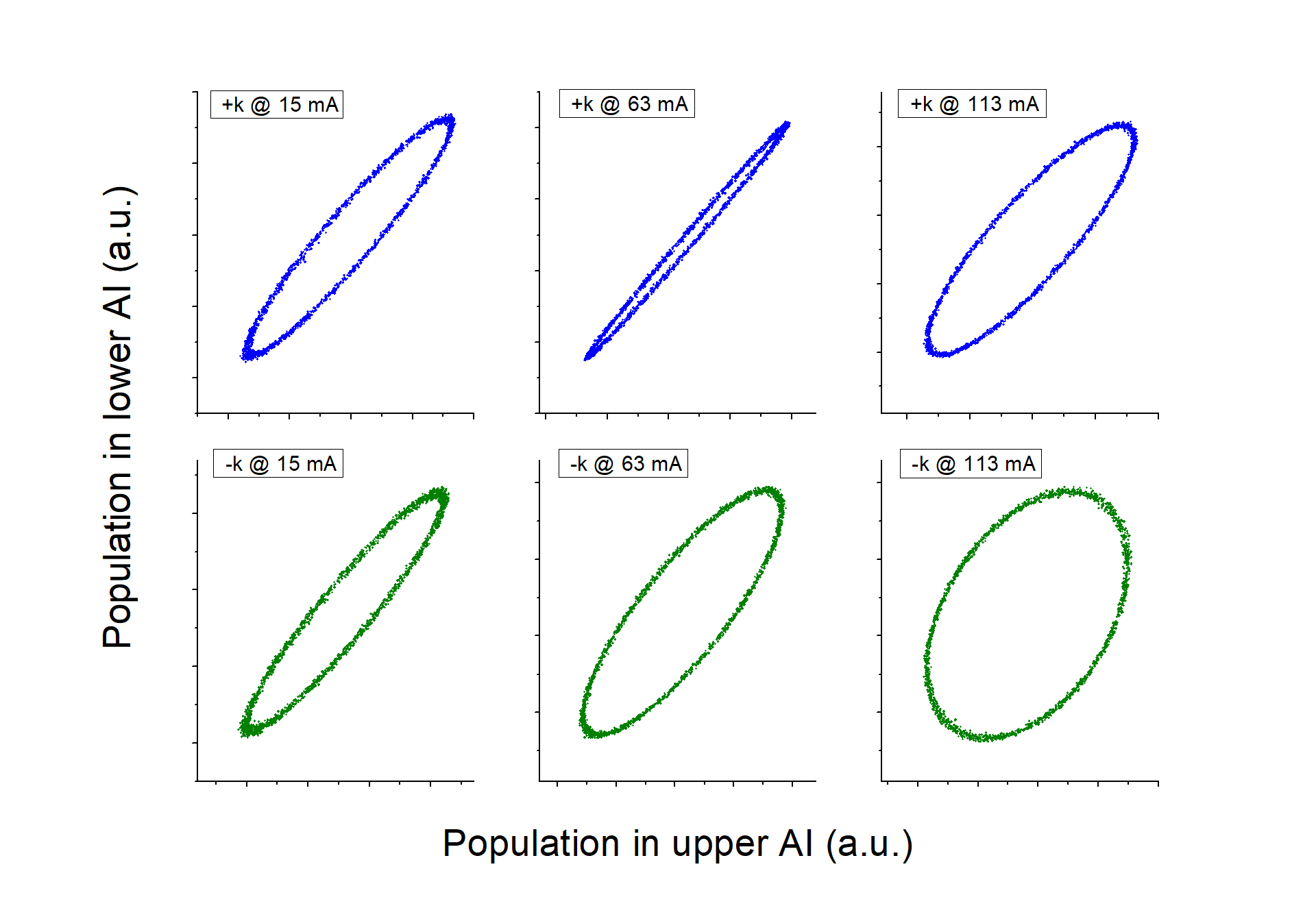}
\caption{\label{8} Ellipses in different magnetic fields. Since ellipse fitting can only give results within [0, $\pi$], we need to combine theoretical prediction to distinguish the sign of the phase, e.g., from the evolution trend of the +k ellipse shape, we can conclude that the phase has undergone a polarity reversal.}
\end{figure}

\begin{figure}[h]
\centering
\includegraphics[width=90mm]{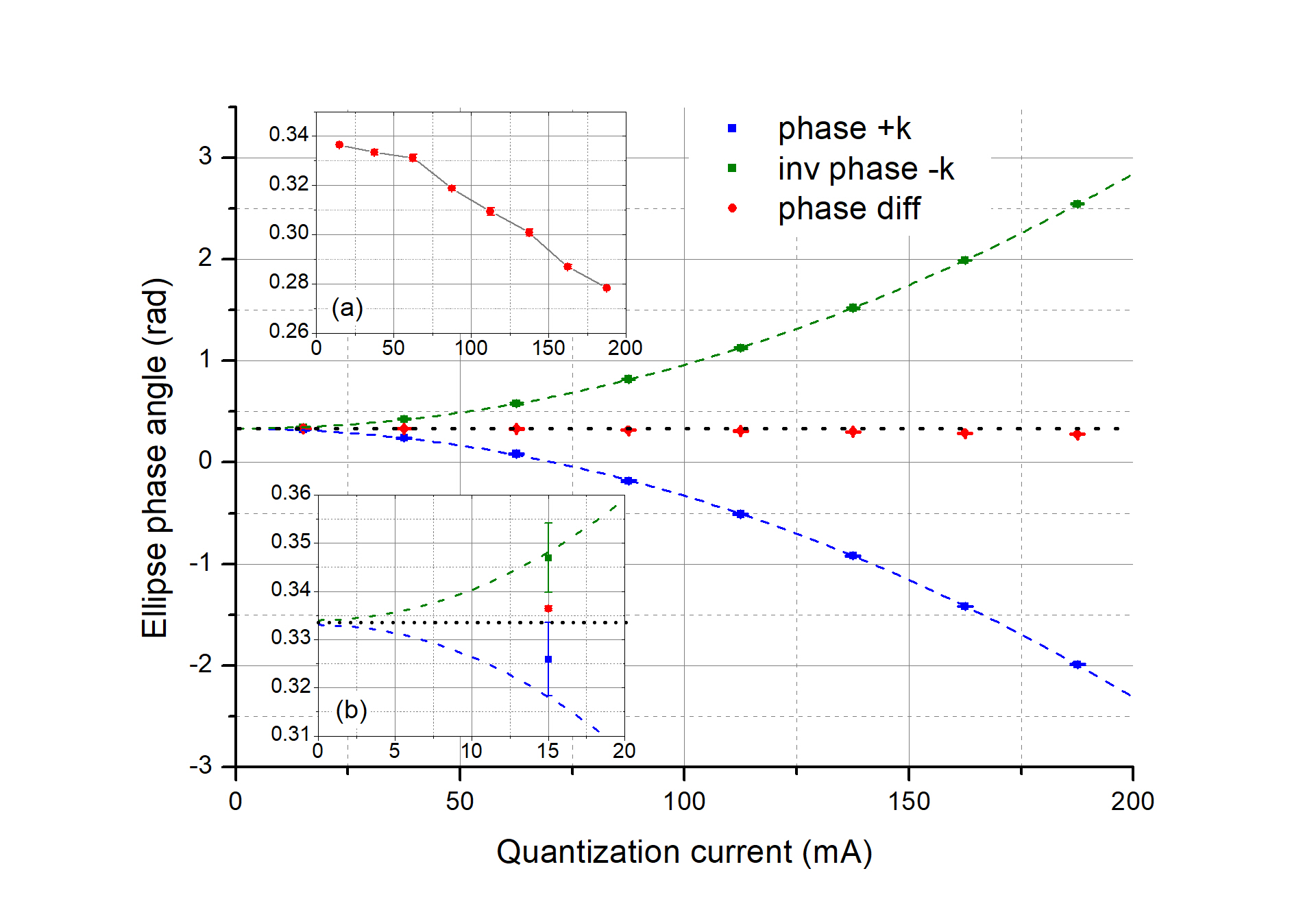}
\caption{\label{9} Dependence of the ellipse phase on the quantization magnetic field. The magnetic fields are scanned three times, within each scan the measurements of 8$\times$40 minutes are taken at different magnetic fields. The dashed lines are parabola-fitted results for +k and -k modes respectively. Insets (a) and (b) are the corresponding enlarged parts of the main curve.}
\end{figure}

The error caused by Zeeman effect has already been studied in Ref. \cite{20}. In this work, as described in sec. IV, we perform an algorithm relying on inversion of the Raman wave vector, thus this effect should have been eliminated in principle. However, when we scan the quantization magnetic fields, as shown in Fig. \ref{8}. and Fig. \ref{9}., though the phase shifts are significantly attenuated by the algorithm, there remains an obvious relationship between the differential phase $(\Delta\varphi_{+k}-\Delta\varphi_{-k})/2$ and the magnetic fields, which is shown in the inset Fig. \ref{9}(a). This is probably caused by the difference between the atom trajectories in the two measurement modes. Since the atoms obtain a recoil velocity of $\sim$12 mm/s from a Raman laser pulse, there is a $\sim$3 mm maximum offset of the average position of atoms between the $+k$ and $-k$ interferometers, that induces different experiences of atoms in non-uniform magnetic fields. Finally, we derive the corrected phase value by extrapolating the two parabola-fitted curves to zero fields. The results from $+k$ and $-k$ interferometers are highly consistent within 1 mrad, as shown in the inset Fig. \ref{9}(b). We thus evaluate a Zeeman correction phase of -3.0±1.0 mrad, which corresponds to a gravity gradient offset of -25±8 E. 


Coriolis effect is a very important error source for atom gravimeters. The phase shift is given by
\begin{equation}
\Delta\varphi_{Coriolis}=\bm{k}_{\mathrm{eff}}\cdot(\bm{\Omega}_{\mathrm{earth}}\times \bm{v})T^2,
\end{equation}
where $\Omega_\mathrm{earth}$ is the earth’s rotation angular velocity, $\bm{v}$ is the average velocity of atoms. Since the wave vector keff is vertical and the earth rotates along the axis pointing north, the phase is sensitive to the velocity component along the east-west direction. Many imperfect experimental conditions, such as misalignment and the power imbalance of the cooling laser beams, will lead to non-zero average horizontal velocity of atoms. For juggling type atom gravity gradiometer, where the two atom clouds are launched by a same group of lasers and tend to acquire similar horizontal velocities, this Coriolis phase shift can be significantly canceled. However, in our gravity gradiometer, the two atom clouds come from different MOTs and probably have different horizontal velocities, thus we have to take this effect as seriously as in atom gravimeters. Although synchronously rotating the Raman laser mirror, that orients $\bm{k}_{\mathrm{eff}}$, is the most popular method to eliminate this phase shift, it is not necessary for a mobile instrument. Here we employ a method of interleaving the orientation of the sensor head to evaluate the Coriolis error. 

The ellipse phase shifts for an atom gravity gradiometer with opposite horizontal orientations, 0° and 180°,  can be written as
\begin{subequations}
\begin{eqnarray}
\Delta\varphi_{0}&&=\Delta\varphi_{gg}+\Delta\varphi_{\mathrm{Coriolis\_0}}\nonumber\\
&&=\bm{k}_{\mathrm{eff}}\cdot\Delta\bm{g}T^2+\bm{k}_{\mathrm{eff}}\cdot(\bm{\Omega}_{\mathrm{earth}}\times\Delta\bm{v}_{\mathrm{h}})T^2,
\end{eqnarray}
\begin{eqnarray}
\Delta\varphi_{180}&&=\Delta\varphi_{gg}+\Delta\varphi_{\mathrm{Coriolis\_180}}\nonumber\\
&&=\bm{k}_{\mathrm{eff}}\cdot\Delta\bm{g}T^2-\bm{k}_{\mathrm{eff}}\cdot(\bm{\Omega}_{\mathrm{earth}}\times\Delta\bm{v}_{\mathrm{h}})T^2,
\end{eqnarray}
\end{subequations}

where the phase shift induced by the gravity gradient $\Delta\varphi_{{gg}}$ remains the same while that induced by the earth’s rotation is reversed in sign. Therefore, the Coriolis phase shift can be evaluated by
\begin{equation}
\Delta\varphi_{\mathrm{Coriolis}}=(\Delta\varphi_{0}-\Delta\varphi_{180})/2,
\end{equation}

and the measurement result with the Coriolis effect eliminated is
\begin{equation}
\Delta\varphi_{gg}=(\Delta\varphi_{0}+\Delta\varphi_{180})/2,
\end{equation}

\begin{figure}[t]
\centering
\includegraphics[width=90mm]{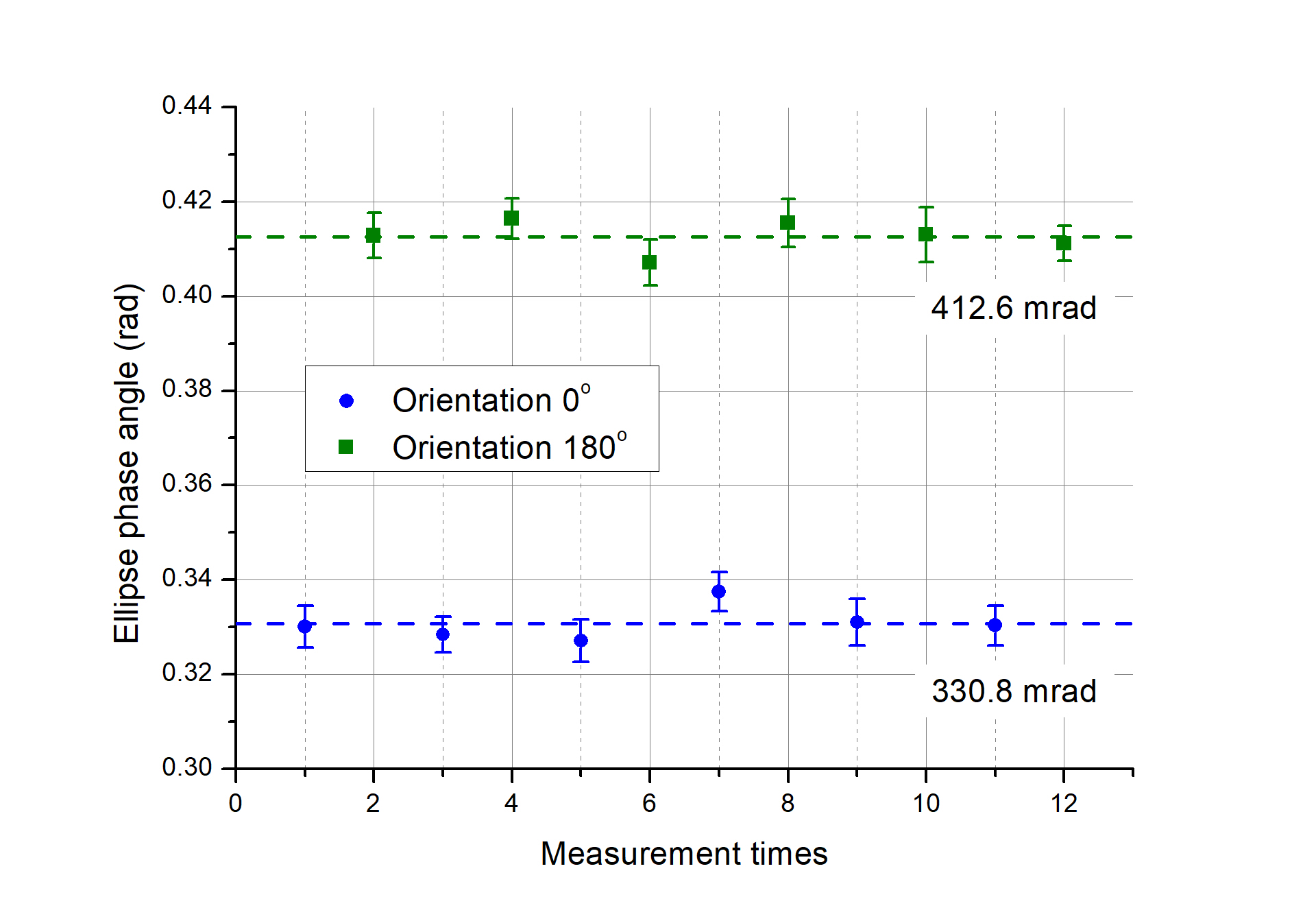}
\caption{\label{10} Dependance of the ellipse phase angle on the horizontal orientation of the sensor head. Each measurement takes 30 minutes.}
\end{figure}

The experimental results are shown in Fig. \ref{10}., each point takes 30 minutes and the phases in two situations are stable over 12 times of alternated measurements. Finally, we derive the corrected phase shift $\Delta\varphi_{{gg}}$=371.7±4.9 mrad, and the phase offset induced by Coriolis effect $\Delta\varphi_{\mathrm{Coriolis}}$=+40.9±4.9 mrad, which corresponds to a gravity gradient of 338±40 E. 

\begin{figure}[h]
\centering
\includegraphics[width=85mm]{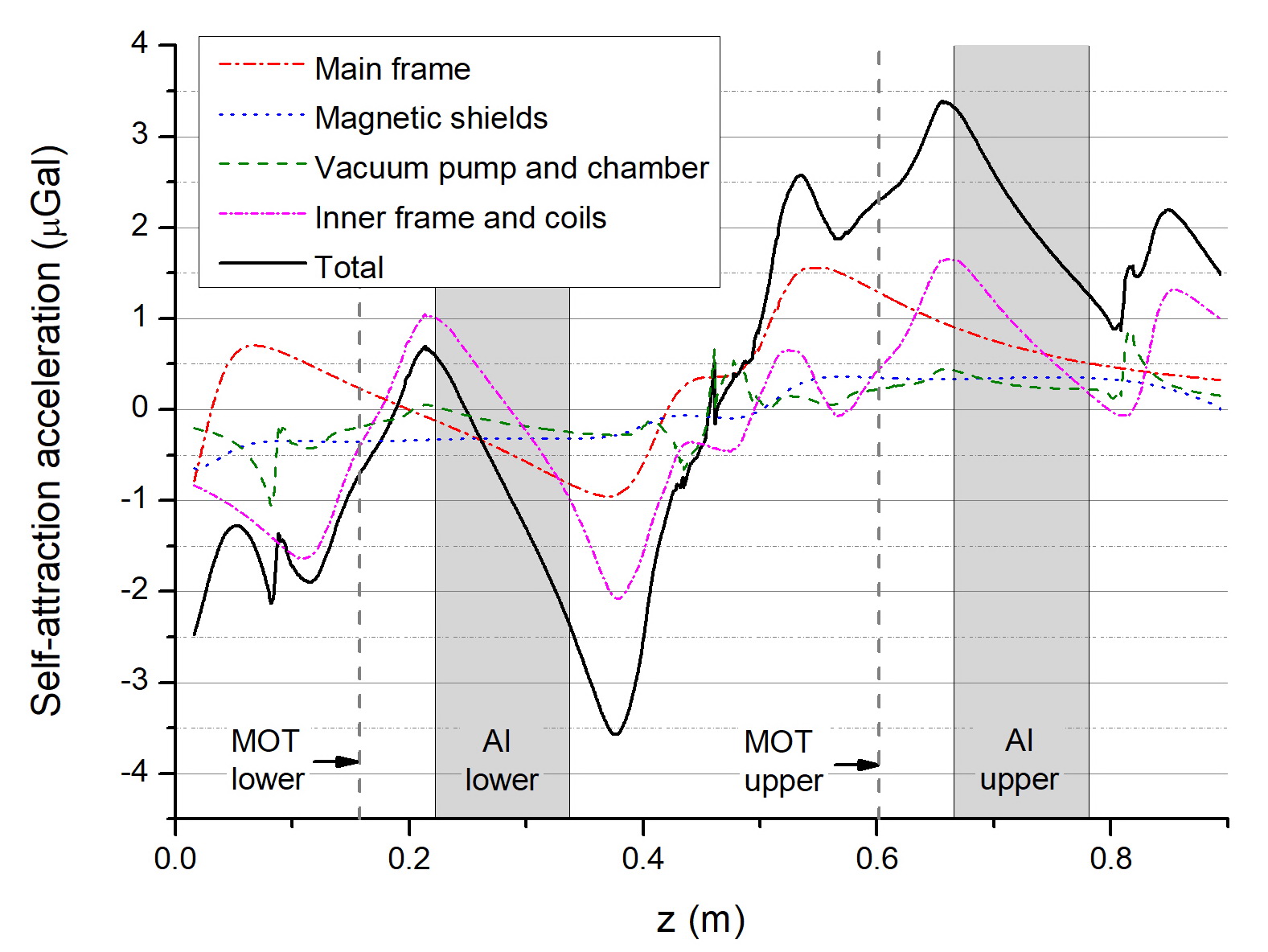}
\caption{\label{11} Self-attraction acceleration along the vertical direction caused by different part of the sensor head. The vertical dashed lines indicate the positions of the two MOTs and the grey shadings indicate the interference regions. The position $z$=0 corresponds to the bottom of the sensor head.}
\end{figure}

Self-attraction effect is another important error source in both atom gravimeters and gravity gradiometers \cite{29,30}. Here we map the self-attraction field by the finite element method, and the induced phase shifts is computed through the perturbation path integral treatment. The self-attraction fields induced by different parts of the sensor head are shown in Fig.\ref{11}. Unlike those previous apparatus where most of the masses distributes on one side of the interference region and give relatively smooth curves, in our gravity gradiometer with two discrete sets of 2D, 3D-MOTs, interference region and a common pump unit in the middle of the two interferometers, the mass distribution is much more scrambled and leads to fluctuant curves. Among those attractive sources, though the magnetic shields account for more than one-third of the total mass, they contribute little to the total attractive field because their mass is evenly distributed around the interferometers. Instead, the frames and coils play the most important roles due to their close distance to the interference regions. The induced phase shift in each interferometer is given by
\begin{eqnarray}
\Delta\varphi&&=\frac{1}{\hbar}\int_{0}^{2T}(L_{pert}^{I}[z(t),\dot{z}(t)]-L_{pert}^{II}[z(t),\dot{z}(t)])\text{d}t\nonumber\\
&&=-k_{\mathrm{eff}}(\int_{0}^{T}ta_{pert}(t)\text{d}t+\int_{T}^{2T}(2T-t)a_{pert}(t)\text{d}t),\nonumber\\
\ 
\end{eqnarray}
where the $L_{pert}^{I}$ are the perturbed Lagrangian along the interferometer path $I$ and $II$, $a_{pert}(t)$ is the perturbed acceleration along the $z$-axis (vertical direction). Then we get the offset of the measured $g$ induced by the self-attraction
\begin{eqnarray}
\delta g=\frac{1}{T^2}(\int_{0}^{T}ta_{pert}(t)\text{d}t+\int_{T}^{2T}(2T-t)a_{pert}(t)\text{d}t).\nonumber\\
\ 
\end{eqnarray}

Combining with atom trajectories, we get the offset of the measured differential gravity $\delta g=\delta g_1-\delta g_2=3.43\ \mu$Gal, which corresponds to gravity gradient of -77 E given the baseline of 44.5 cm. Although the trajectory deflections induced by the photon recoil momentums are also considered in our calculation, the difference between the $+k$ and $-k$ measurements is found to be only 0.3 E. Taking into account the neglected small components, non-uniform density distribution of commercial devices, and non-zero atom cloud sizes, etc., we give an uncertainty of this offset as ±5 E.  

The systematic errors and the uncertainties are summarized in Table 1, where the Coriolis effect gives the dominant offset and uncertainty. In principle, the error caused by Raman wavefront distortion, which usually contributes largest error and uncertainty in atom gravimeters, could be common-mode rejected suppose the two atom clouds are perfectly overlapped and expand synchronously. However, that ideal situation will not exist, thus we estimate this uncertainty to be 33 E under the condition that the mirror and the vacuum window distort the wavefront to $\lambda/5$ and the two atom clouds horizontally offset by 1 mm. From the budget above, we measured the local gravity gradient and obtained the corrected result of 3114±53 E, which agrees with the mean free-air value of 3090 E \cite{31}. 

\begin{table}[h]
\caption{\label{table1}Systematic error budget}
\begin{ruledtabular}
\begin{tabular}{lcc}
\textrm{Systematic effect}&
\textrm{Correction (E)}&
\textrm{Uncertainty (E)}\\
\colrule
Laser frequency & 0 & $<0.1$\\
Vibration & 0 & $<0.1$\\
Vertical alignment & 0 & $<0.1$\\
Baseline & 0 & 4\\
Raman wavefronts & 0 & 33\\
Ellipse fitting & -14 & 5\\
AC Stark shift & +91 & 6\\
Magnetic effect & -25 & 8\\
Coriolis effect & +338 & 40\\
Self-attraction & +77 & 5\\
\colrule
\rule{0pt}{12pt}
$\bm{\mathrm{Total}}$ & $\bm{\mathrm{+467}}$ & $\bm{\mathrm{53}}$\\
\end{tabular}
\end{ruledtabular}
\end{table}

\section{\label{sec:level1}Conclusion and discussion}

We have developed a compact high-resolution gravity gradiometer based on dual Rb-85 LPAIs for the measurement of vertical gradient component. Two novel all-glass vacuum chambers were designed and fabricated, which suppress the volume of the sensor head down to an unprecedented 95 liters. The weight of the sensor head is only 66.4 kg and can be easily carried by two people. In addition to the optimization of many experimental parameters, the wave-vector-inversion algorithm was implemented and demonstrates remarkable power of long-term drift suppression, that leads to a measurement resolution up to 0.86 E after an integration of 17000 s. Therefore, it is the sub-E atom gravity gradiometer with highest level of compactness to date as far as we know. Since the previous most compact atom gravity gradiometer has a sensor head’s volume of more than 200 liters \cite{19}, this 95 liters sensor head will bring great convenience in matching with inertial platforms and demonstrate significant advantage in carriers where space is at a premium. The systematic errors were carefully evaluated. A special error, ac-Stark shift imbalance, caused by Raman laser scheme based on chirping AOM, was found and eliminated. We measured the local absolute gravity gradient with this gravity gradiometer and obtained a result of 3114±53 E, which agrees well with the reference mean free-air value. 

The performance of the gravity gradiometer can be further improved in the future. The weight of the sensor head can be easily decreased to less than 50 kg by optimization of the supporting structure, which will further reduce the requirement to the stabilized platform. The detection noise is probably the limit to the short-term sensitivity, it can be decreased by active control of the detection laser’s intensity and utilization of detectors with lower noises. The simultaneous detection of two atomic states \cite{27}, which is insensitive to the fluctuations of laser frequency and intensity, is also a scheme very worth trying. The residual long-term drift is mainly caused by the temperature effect of the optical fibers, thermal feedback control to critical devices will be helpful for improving the long-term stability. With respect to the accuracy, the Coriolis error could be evaluated more accurately by using a precision rotating platform, and the uncertainty caused by the effect of wavefront distortion could be reduced through the extrapolation experiment of beam size modulation \cite{32}.

\begin{acknowledgments}
We acknowledge the financial support from the National Key search and Development Program of China under Grant No. 2016YFA0302002, the National Natural Science Foundation of China under Grant No. 91536221, and No. 91736311, the Strategic Priority Research Program of Chinese Academy of Sciences under Grant No. XDB21010100.
\end{acknowledgments}

\nocite{*}

\bibliography{vgg}

\end{document}